\documentclass[doublecol,linenumbers]{epl2} 

\title{Improving performance of inverse Compton sources through laser chirping}
\shorttitle{Improving performance of inverse Compton sources through laser chirping} 

\author{B. Terzi{\'c}\inst{1} \and G. A. Krafft\inst{1,2} \and A. Brown\inst{1} \and
T. Hagerman\inst{1} \and E. Johnson\inst{1} \and V. Petrillo\inst{3,4} \and
I. Drebot\inst{3} \and C. Maroli\inst{3} \and M. Ruijter\inst{3,5}
}
\shortauthor{B. Terzi{\'c} \etal}

\institute{                    
  \inst{1} Department of Physics, Center for Accelerator Science, Old Dominion University, Norfolk, Virginia 23529, USA \\
  \inst{2} Thomas Jefferson National Accelerator Facility, Newport News, Virginia 23606, USA \\
  \inst{3} INFN-Milan, via Celoria 16, 20133 Milano, Italy \\
  \inst{4} Universita degli Studi di Milano, via Celoria 16, 20133 Milano, Italy \\
  \inst{5} Universita di Roma La Sapienza, P.~le A.~Moro 5, 00185 Roma, Italy \\  
  }

\abstract{
We present a new paradigm for computation of radiation spectra in the non-linear regime of operation
of inverse Compton sources characterized by high laser intensities. The resulting 
simulations show an unprecedented level of agreement with the experiments. Increasing the laser 
intensity changes the longitudinal velocity of the electrons during their collision, leading to 
considerable non-linear broadening in the scattered radiation spectra. The effects of such 
ponderomotive broadening are so deleterious that most inverse Compton sources either 
remain at low laser intensities or pay a steep price to operate at a small fraction of the physically 
possible peak spectral output. This ponderomotive broadening can be reduced by a suitable 
frequency modulation (also referred to as ``chirping'', which is not necessarily linear) of the 
incident laser pulse, thereby drastically increasing the peak spectral density. This frequency 
modulation, included in the new code as an optional functionality, is used in simulations to 
motivate the experimental implementation of this transformative technique.
}

\begin{document}
\maketitle

\section{Introduction}
When a relativistic electron beam interacts with a high-field laser beam, intense and highly 
collimated electromagnetic radiation will be generated through Compton scattering 
\cite{jackson,priebe}. 
Through relativistic upshifting and the relativistic Doppler effect, 
highly energetic polarized photons are radiated along the electron beam motion when the 
electrons interact with the laser light. For example, x-ray radiation can be obtained when 
optical lasers are scattered from electrons of tens-of-MeV beam energy. Because of the 
desirable properties of the radiation produced, many groups around the world have been 
designing, building, and utilizing inverse Compton sources (ICS) for a
wide variety of purposes. Sources of electromagnetic radiation relying upon 
Compton scattering are being applied in fundamental physics research and compact 
accelerator-based sources specifically designed for potential user facilities have been built 
\cite{huang}. One remarkable feature of the radiation emerging from such sources, 
compared to bremsstrahlung sources, is its narrow-band nature. 
Applications to x-ray structure determination \cite{aetal2010}, 
dark-field imaging \cite{betal2009,setal2012}, phase contrast imaging \cite{betal2009}, and 
computed tomography \cite{aetal2013} have been demonstrated experimentally and take full advantage of the narrow bandwidth of ICS.

\begin{figure}[htb]
\vskip-20pt 
\vskip-0.25in
\includegraphics[width=3.1in,height=3.45in,angle=270]{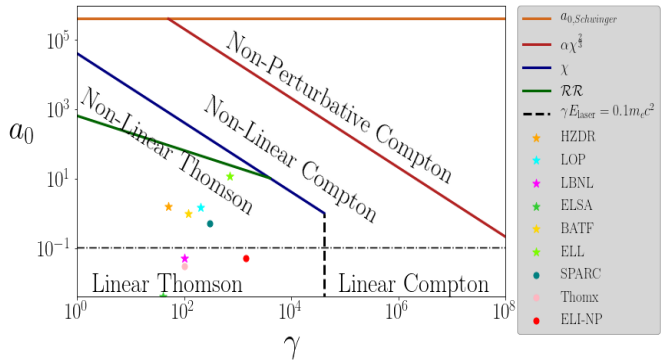}
\vskip-43pt
\caption{\small{Regimes of operation for ICS.
Red line denotes the boundary between the non-perturbative and perturbative
non-linear Compton regimes \cite{nr1964,f2017}. 
Blue line indicates the division between Thomson and perturbative Compton scattering. 
Between the blue and green line radiation reaction needs to be taken into account  
\cite{d2008,hetal2010,retal2018}.
ICS that are in operation are marked as stars and future ICS as dots   
\cite{HZDR,LOP,BATF,LBNL,ELSA,SPARC,Thomx}.
}}
\label{fig_regimes}
\end{figure}

Depending on the properties of the two fundamental elements of 
ICS---the energy of an electron beam and the intensity of a laser---there are several regimes 
of operations, shown in Fig.~\ref{fig_regimes}. 
With the increasing electron beam energies, there are:
(i) Thomson regime at low-to-medium electron beam energies
($2\gamma E_{\rm laser} \ll m_e c^2$), where the electron recoil can be neglected; and 
(ii) Compton regime at high electron beam energies
($2\gamma E_{\rm laser} \sim m_e c^2$), requiring proper accounting for electron recoil.
As the laser intensity increases, there are: (i) linear regime at low 
laser intensities; and (ii) non-linear regime at high laser intensities. 
The onset of non-linearity is quantified by the increase in the amplitude 
of the normalized vector potential ($a_0$) representing the laser 
${\tilde A}(\xi)=eA(\xi)/m_e c =a(\xi) \cos (2\pi \xi/\lambda_0)$, where $a(\xi)$ is the laser 
envelope, $\xi=z+ct$ the coordinate along the laser pulse and $\lambda_0$ the mean
wavelength of the laser.

In this letter, we develop a paradigm for computation of radiation spectra 
emitted from ICS operating in the non-linear Thomson regime, extending  to 
high laser intensities and low-to-medium electron beam energies.
The resulting new computer code, SENSE (Simulation of Emitted Non-linear Scattering Events), 
uses a three-dimensional (3D) 
pulse model for the laser beam, a significant generalization of the one-dimensional (1D) 
plane-wave model. The electron beam is either generated by random sampling its bulk 
properties or supplied as input.

\section{SENSE code}

SENSE computes spectra of the scattered radiation in 
ICS by integrating a spectrum $d^2 E/d\omega d\Omega$ due to a collision
of a single electron with a 3D laser pulse over an entire distribution of electrons. 
Monte Carlo integration over a solid angle $d\Omega(\theta, \phi)$ 
of the physical aperture with the angular size of $\theta_a$ is used to compute a spectrum 
${{d E(\omega)}/{d\omega}} = \int {{d^2 E(\omega; \Omega)}/({d\omega d\Omega}}) d\Omega$
for each of $N_s$ simulation particle sampling a distribution of $N_e$ electrons.
The total spectrum is the average of these individual spectra.
It is written in Python, and uses Cython 
and {\tt numpy} for computational efficiency \cite{python,cython}. It is parallelized to 
run on multiple CPUs.

\begin{figure}[htb]
\includegraphics[width=3.45in]{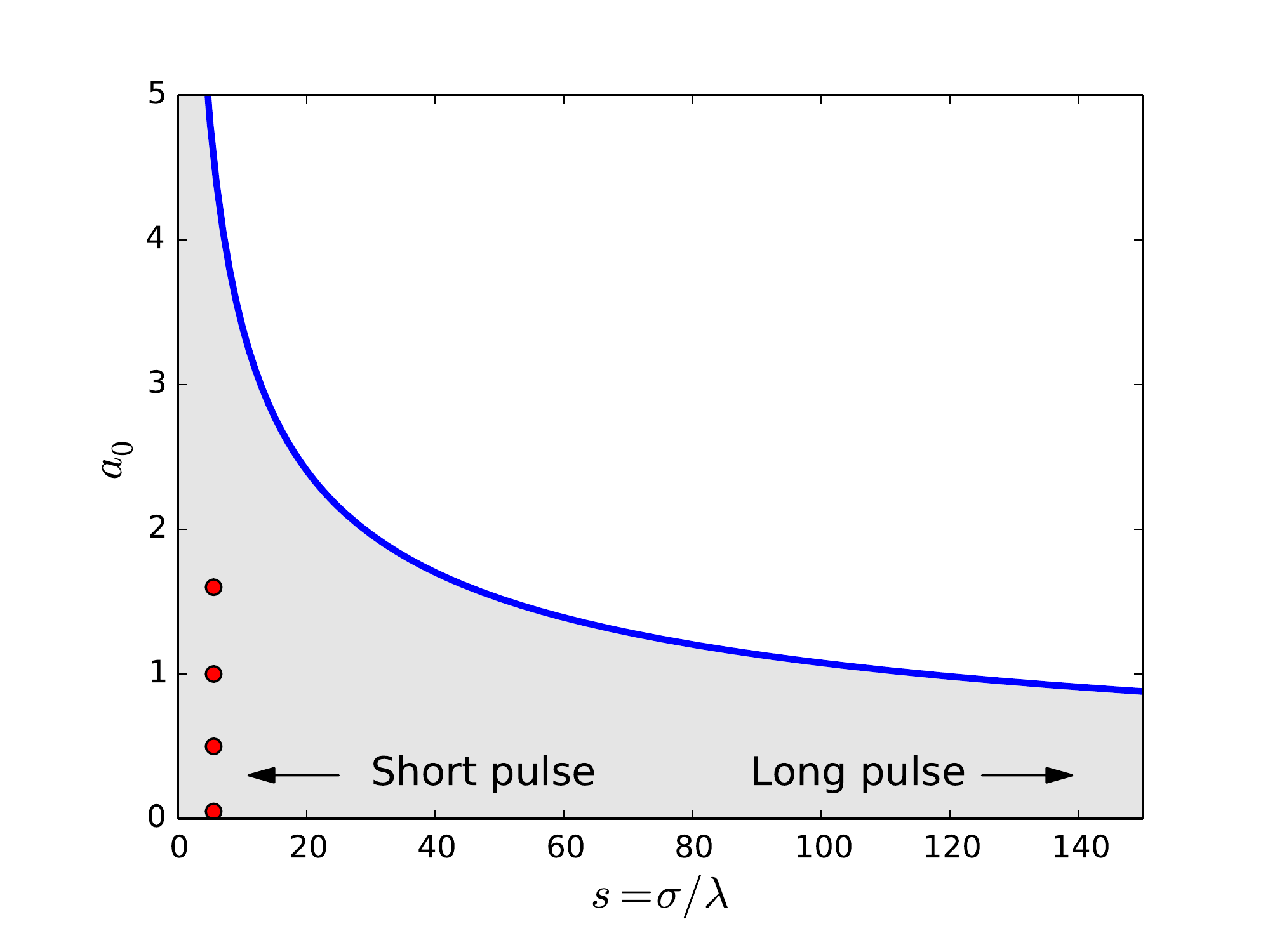}
\caption{\small{Region of validity for SENSE shown in gray. Red dots denote the four
experimental values for the laser field strength $a_0$ from the Dresden experiment \cite{ketal2018}.
}}
\label{fig_validity}
\end{figure}

SENSE is applicable in the non-linear Thomson regime, where electron recoil can be neglected. 
The range of the laser
field parameter $a_0$ for which this formalism is applicable is derived by requiring that
the total number of photons emitted is less than one \cite{PRL}, and is given by 
$a_0 < \sqrt{3\lambda_0/(2\pi^{1/2} \alpha \sigma_{l,z})}$, with $\sigma_{l,z}$ the laser pulse
length and $\alpha$ the fine structure constant.
For the Dresden experiment \cite{ketal2018} simulated in the next section 
($s = \sigma_{l,z}/\lambda_0 = 5.57$, so SENSE is applicable for
the field strength parameters $a_0 < 4.56$. All simulations reported here are well within
this limit, as are most other existing and future ICS. The parameter space for which
SENSE is applicable is shown in Fig.~\ref{fig_validity}.

There is a fundamental difference between the Monte Carlo implementation in CAIN 
\cite{CAIN} and SENSE. Both codes start by randomly sampling the electron 
distribution. However, CAIN models the incoming laser beam scattering off each such electron 
from the sample with a number of individual scattered particles. While this directly models what happens in an 
experiment, albeit on orders-of-magnitude smaller scales, it ensures that the rare events in 
nature will be equally rare in a simulation, leading to poor statistics in those regions. 
In contrast, SENSE computes scattering probabilities---the likelihood of scattered 
photons to be found in each portion of the spectrum. Therefore, the accuracy in 
each portion of the spectrum computed by SENSE
is the same, determined only by the accuracy of the Monte Carlo integration.

There are two important features of SENSE that CAIN either does not have or implements 
only in a cumbersome way: 
(1) an arbitrary shape of the laser pulse and 
(2) an arbitrary laser frequency modulation (FM) scheme. 

SENSE uses the code designed in \cite{k2004} and amended 
for FM in \cite{PRL} for backscattered, on-axis photons ($\phi=0$, $\theta=0$). 
It is first generalized to arbitrary angles $\phi$ and $\theta$, in order to 
evaluate the total scattered radiation spectrum:
\begin{equation}
{{d^2 E}\over{d\omega d\Omega}} = {{d^2 E_\sigma}\over{d\omega d\Omega}} 
+ {{d^2 E_\pi}\over{d\omega d\Omega}},
\end{equation}
where
\begin{eqnarray}
{{d^2 E_\sigma}\over{d\omega d\Omega}} & = & {{e^2}\over{8 \pi^2 c^3}} 
\omega^2 \left|D_x\right|^2 \sin^2 \phi, \\
{{d^2 E_\pi}\over{d\omega d\Omega}} & = & {{e^2 }\over{8 \pi^2 c^3}} \omega^2 \left|D_x \left({{\cos\theta - \beta_z}\over{1-\beta_z\cos\theta}}\right) 
\cos\phi + D_z\sin \theta\right|^2,  \nonumber
\end{eqnarray}
\begin{widetext}
\begin{equation} \label{eq:Dx}
D_{x,z} = c_{x,z}  \int^\infty_{-\infty} \tilde{A}^{1,2}(\xi) d\xi 
 \exp\left[i\omega\left({{\xi(1-\beta_z\cos\theta)}\over{c(1+\beta_z)}}
-{{\sin\theta \cos\phi}\over{c\gamma(1+\beta_z)}}\int_{-\infty}^\xi\tilde{A}(\xi')d\xi'
+{{(1+\cos\theta)}\over{2c\gamma^2(1+\beta_z)^2}}\int_{-\infty}^\xi\tilde{A}^2(\xi')d\xi'
\right)\right]. 
\end{equation}
\end{widetext}
SENSE models electron beam's emittance and the energy spread  
with a geometric argument. The 3D pulsed nature of a laser is modeled by 
varying effective field parameter for each electron based on its
path through the laser. 

An electron along the $z$-axis of collision passes through the laser pulse head on. 
An electron with transverse motion 
$p_x$, $p_y\ne 0$, will pass through the laser pulse at an angle, thereby extending its path by
$1/r$:
\begin{equation} \label{eq_r}
r \equiv (p_z/\gamma)/\sqrt{p_x^2 + p_y^2 + (p_z/\gamma)^2} \le 1.
\end{equation}
Because each electron passing through a laser pulse see the same number of wavelengths, 
extending the path traveled means that the effective wavelength of the laser is increased
$\lambda_0 = {{{\bar \lambda}_0}/{r}}$,
or, equivalently, the frequency is decreased (``red-shifted'') $\omega_0 = r {\bar \omega}_0$.
Barred quantities are experimental parameters.

\begin{figure}[htb]
\vskip-10pt
\includegraphics[width=3.45in]{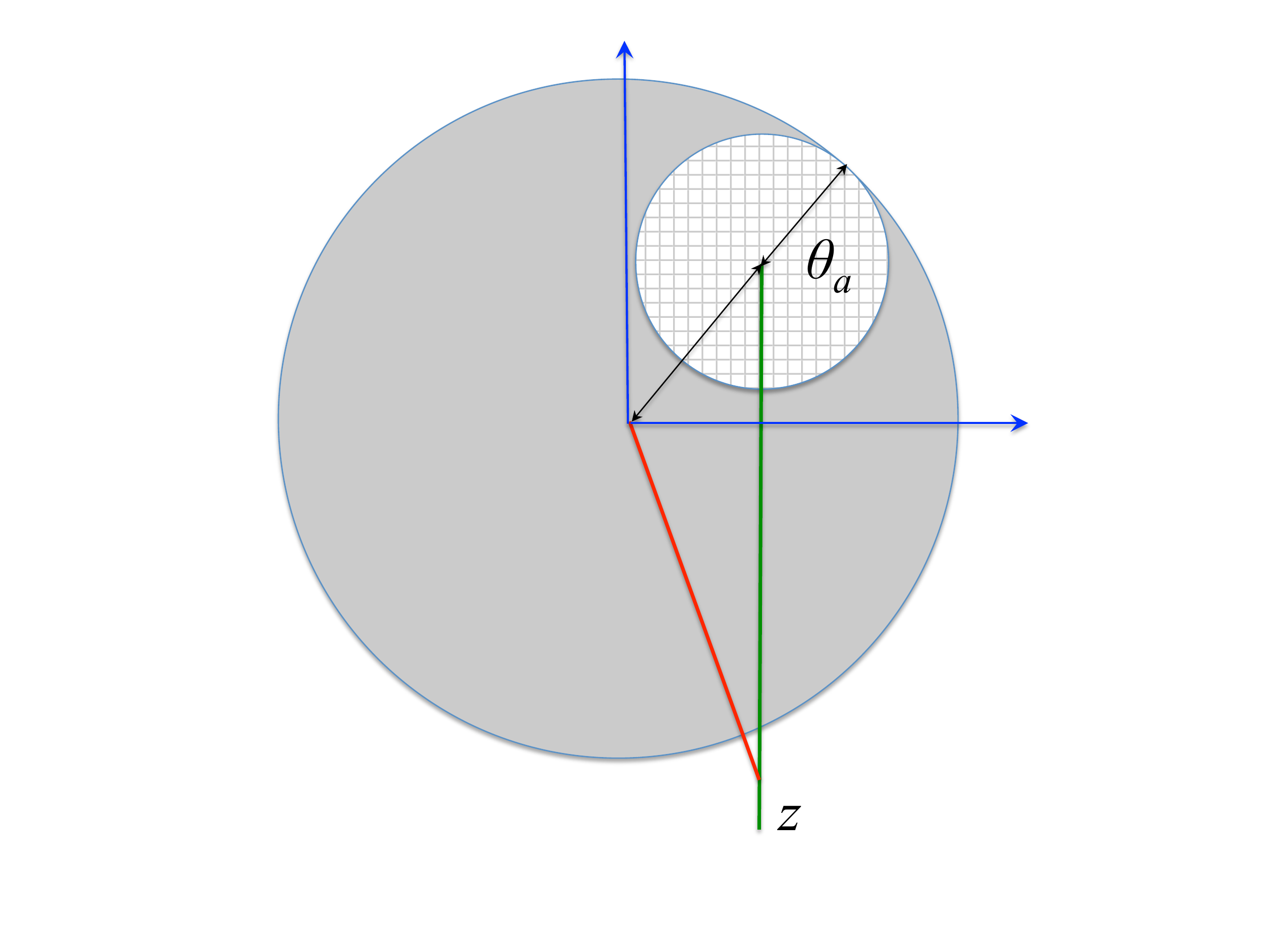}
\vskip-25pt
\caption{\small{
Geometry of photon scattering. Checkered region represents physical aperture $\theta_a$. 
Shaded region is centered on the point where a scattered photon pierces the plane of the 
physical aperture. Green line is the $z$-axis. Red line denotes the path of a 
scattered photon. 
Polar coordinates $(\cos \theta, \phi)$ in which the integration is carried out are shown in blue. 
}}
\label{fig1}
\end{figure}

The resulting effects on scattered radiation frequency are obtained from 
$\omega = (1+\beta)^2 \gamma^2 \omega_0$. The effects of the energy spread can 
be found by replacing 
$\gamma \approx  {\bar \gamma} \left(1 + {{\Delta \gamma}/{\bar \gamma}} \right)$
to obtain
$
\omega = {\bar \omega} r \left(1+2 {{\Delta \gamma}/{\bar \gamma}}\right) \equiv k {\bar \omega},
$
where $k \equiv r \left(1+2 {{\Delta \gamma}/{\bar \gamma}}\right)$.
This means that in order to properly account for angles (emittance) and the energy spread, 
the computed spectra should be red-shifted by a factor $1/k$.
Therefore, SENSE computes
\begin{equation} \label{eq_spectrum_MCb}
\left({{d E(\omega)}\over{d\omega}}\right)_{\rm beam} =
{{N_e}\over{N_s}} \sum_{i=1}^{N_s}
{{d E(\omega/k(x,y,p_x,p_y,\gamma))}\over{d\omega}}.
\end{equation}
$dE/d\omega$ is computed by integrating $d^2E/(d\omega d\Omega)$ over
the physical aperture as shown in Fig.~\ref{fig1}.

In the 3D laser pulse model, the strength of the effective laser field that an electron 
experiences depends on its path. An on-axis electron ``sees'' the maximum strength.

We model electrons with angles as taking straight paths through the laser pulse.
For an electron at the beginning of interaction with 
the laser beam ($t = 0$), the spatial coordinates are $(x_0, y_0, z_0)$. 
The time coordinate is $t = (\xi - z_0)/c$. 
After normalizing transverse momenta
${\tilde p}_{x,y} \equiv {{p_{x,y}}/{(m_e c)}}$, an electron's trajectory is:
\begin{equation}
x = x_0 + {\tilde p}_{x} \xi, \hskip10pt y = y_0 + {\tilde p}_{y} \xi, \hskip10pt  z = z_0 + r \xi.
\end{equation}
In these new coordinates, a gaussian laser pulse with rms size
$\sigma_{x,l}$, $\sigma_{y,l}$, $\sigma_{z,l}$ as experienced by the electron is
\begin{equation} \label{eq_pulse_2a}
a (\xi) =
{\tilde a}_0 
\exp\left(-{{(\xi + \eta)^2}/{(2{\tilde \sigma}_{z,l}^2)}}\right), 
\end{equation}
where
\begin{eqnarray}
{\tilde a}_0 & = & a_0 \exp\left( -{{x_0^2}\over{2\sigma_{x,l}^2}} -{{y_0^2}\over{2\sigma_{y,l}^2}}
- {{z_0^2}\over{2\sigma_{z,l}^2}}\right)
\exp\left({{\eta^2}\over{2{\tilde \sigma}_{z,l}^2}}\right), \nonumber \\
\eta & = & {\tilde \sigma}_{z,l}^2 \left({{x_0{\tilde p}_{x}}\over{\sigma_{x,l}^2}} +{{y_0 {\tilde p}_{y,l}}\over{\sigma_{y,l}^2}} +{{z_0 r}\over{\sigma_{z,l}^2}}\right), \nonumber \\
{\tilde \sigma}_{z,l}^2 & = &  
{{\sigma_z^2}\over{r^2}} 
\left({{{\tilde p}_x^2 \sigma_z^2}\over{r^2 \sigma_{x,l}^2}} + 
{{{\tilde p}_y^2 \sigma_{z,l}^2}\over{r^2 \sigma_{y,l}^2}} +1\right)^{-1}. \nonumber
\end{eqnarray}
The maximum magnitude of the vector potential $a$ occurs at the center  
of the pulse, at $z_0 = 0$.
This means that the new laser pulse shape given in Eq.~(\ref{eq_pulse_2a}) is also a gaussian, 
only with a changed size ${\tilde \sigma}_{z,l}$ and the amplitude of the normalized vector 
potential ${\tilde a}_0$.
The change in the path length of an electron's passage through the laser pulse is 
\begin{equation}
{\tilde r} \equiv r
\sqrt{{{{\tilde p}_x^2 \sigma_{z,l}^2}\over{r^2 \sigma_{x,l}^2}} + 
{{{\tilde p}_y^2 \sigma_{z,l}^2}\over{r^2 \sigma_{y,l}^2}} +1},
\end{equation}
shifting the wavelength $\lambda_0 = {{{\bar \lambda}_0}/{{\tilde r}}}$.
For electrons with angles $p_x, p_z \ne 0$, the ratio $\tilde r$ can be smaller (larger)
than unity, in which case the frequencies are red-(blue-)shifted.

\section{Simulating experimental results} \label{sec:results}

We simulate the Dresden experiment \cite{ketal2018} using both
CAIN \cite{CAIN} and SENSE. We assume that both the laser pulse and the electron beam
are gaussian-distributed with rms sizes as reported in Table \ref{tab0}\cite{ketal2018}.
At the collision, the centers of the two beams overlap.
The simulations with both CAIN and SENSE seem to be insensitive to the size and shape of
the aperture. The aperture used in SENSE simulation is circular, while the physical aperture 
used in the Dresden experiment was rectangular \cite{ketal2018}. In all of our simulations---those
reported here and many others---the agreement between the results produced by CAIN and 
SENSE is remarkable, especially considering that they are based on two vastly 
different approaches.

\begin{figure}[htb]
\includegraphics[width=3.48in,height=5in]{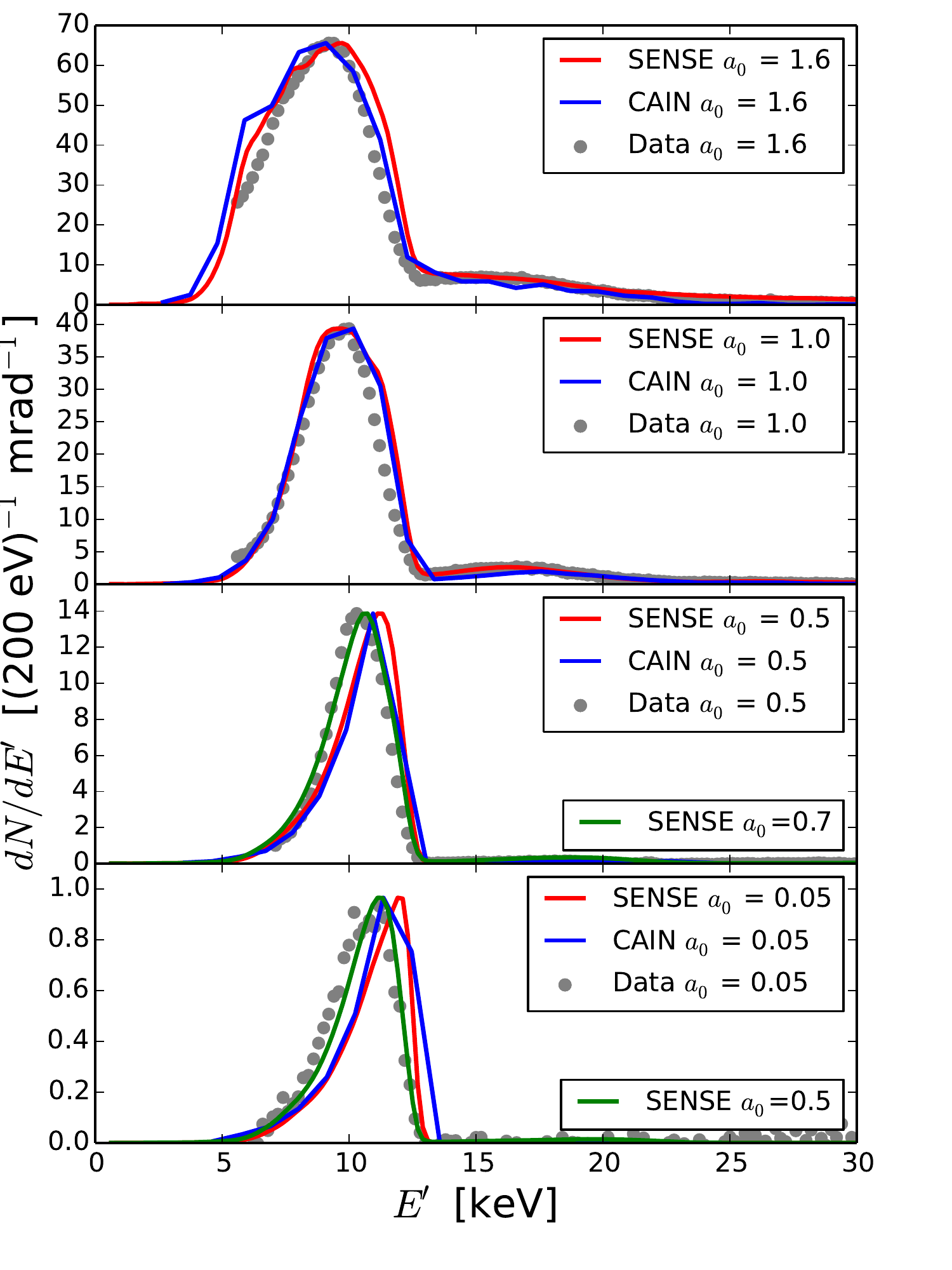}
\vskip-18pt
\caption{\small{Simulation of the Dresden experiment \cite{ketal2018} data (gray circles)
with CAIN (blue line) and with SENSE (red line) for parameters 
$a_0=1.6, 1.0, 0.5, 0.05$. Simulations with SENSE use $N_s$ = 4000 particles and 
circular aperture of $\theta_{\rm a} = 0.004$.
}}
\label{fig_compare}
\end{figure}

The simulations in Fig.~\ref{fig_compare} model the results of Fig.~3 from \cite{ketal2018}. 
For the largest values of the laser field, $a_0 = 1.6, 1.0$, the agreement between the experiments
and simulations using CAIN is very good, and SENSE even better.
However, for lower values, $a_0 = 0.5, 0.05$, there is a shift to the right in the 
simulations from both codes.
Increasing the strength of the laser field from $a_0=0.5$ to $0.7$ and from
$a_0=0.05$ to 0.5 in SENSE simulation produces excellent fits to the data, comparable to 
those for the larger values of $a_0$. The discrepancy between the experiments
and the simulations for the lower values of the strength of the laser field is likely due to a different
geometry of collision. It is unclear which of the geometries reported in Fig.~2 of \cite{ketal2018}
was used in experiments.

\begin{table*}
\begin{center}
\footnotesize
        \setlength\tabcolsep{4pt}
        \begin{tabular}{|l |c |c |c || l |c |c |c|}
                \hline 
                \multicolumn{4}{|c||}{{\bf Laser Pulse}} & \multicolumn{4}{|c|}{
{\bf Electron Beam}} \\
                \hline                
                {\bf Quantity} & {\bf Variable} & {\bf Value} & {\bf Unit} &
                {\bf Quantity} & {\bf Variable} & {\bf Value} & {\bf Unit} \\
                \hline
                Wavelength & $\lambda_0$ & $800$ & nm& Energy & $E_e$ & $23$ & MeV \\
                Pulse duration & $T$ & 14.86 & fs & Energy spread & $\Delta E_e/E_e$ & 0.00175 & \\
                Horizontal spot size & $\sigma_{l,x}$ & 13.59 & $\mu$m & Horizontal spot size & $\sigma_x$ & $41 \pm 1.2$ & $\mu$m \\
                Vertical spot size & $\sigma_{l,y}$ & 13.59 & $\mu$m & Vertical spot size& $\sigma_y$ & $81 \pm 2$ & $\mu$m \\
                Pulse length & $\sigma_{l,z} \equiv cT$ & 4.5 & $\mu$m & Horizontal emittance (normalized) & $\epsilon_{x,n}$ & $20.3 \pm 1.1$ & mm mrad \\
                Normalized length & $s \equiv \sigma_{l,z}/\lambda_0 $ & 5.57 &   & Vertical emittance (normalized) & $\epsilon_{y,n}$ & $18.0 \pm 6.6$ & mm mrad \\
                \hline
        \end{tabular}
        \caption{Laser pulse and electron beam parameters from the Dresden experiment \cite{ketal2018}. All $\sigma$ quantities are reported as rms.}
\label{tab0}
\normalsize
\end{center}
\end{table*}

\section{Improved performance via laser chirping}

In \cite{PRL}, we presented a novel and quite general analysis of the interaction of a 
high-field FM (chirped) 1D plane-wave laser and a relativistic electron, in which 
exquisite control of the spectral brilliance of the up-shifted Compton scattered photon 
is shown to be possible. We showed that the ponderomotive broadening can be eliminated 
by suitable FM of the incident laser. We suggested a practical realization 
of this compensation idea in terms of a chirped-beam-driven free electron laser oscillator 
and showed that significant compensation can occur, even with the imperfect 
matching.

Extending the FM technique from the 1D plane-wave to the 3D pulse model for the laser has 
been carried out recently \cite{metal2018}. Because the electrons colliding with a 3D laser
pulse encounter a full range of laser field strengths $a$, from 0 to $a_0$---depending which portion 
of the pulse they pass through---the $a$-dependent FM of the laser pulse cannot recover the 
narrow bandwidth of every electron in the distribution. Here we seek to answer
by how much can the peak spectral density be increased by a FM of the laser pulse 
and when is FM most effective.

SENSE is capable of simulating FM of any form. We carried out simulations for a 
gaussian laser pulse with three FM prescriptions: 
\begin{enumerate}
\item Optimal chirping for the 3D laser pulse model \cite{metal2018}:
\begin{equation}
f_{\rm 3D}(Y;p) = f_0 \left({{p}\over{3}} + {{1}\over{Y}} \int_0^Y d Y' (s_1'(Y')+ s_2'(Y'))\right),
\nonumber
\end{equation}
where 
$A(Y) = a_0 \exp\left(-2Y^2\right)/2$, $Y = \xi/(\sqrt{2} \sigma)$, and $p$ is an 
arbitrary constant and
\begin{equation}
s_{1,2} = \left[{{p}\over{2}} A(Y) + {{p^3}\over{27}} \pm
\sqrt{
{{p^4}\over{27}} A(Y) + 
{{p^2}\over{4}} A^2(Y) 
}
\right]^{1/3}. \nonumber
\end{equation}
\item Optimal chirping for the 1D plane-wave model \cite{PRL}:
\begin{equation}
f_{\rm 1D}(\xi; a) = f_0 \left(1+{{\sqrt{\pi}\sigma a^2}\over{4\xi}} {\rm erf} ({{\xi}/{\sigma}})\right),
\nonumber
\end{equation}
with $a$ the laser field strength, which varies from 0 to the maximum field strength $a_0$.
\item FM produced by a free electron laser oscillator.
When a driving beam bunch is long enough that the radio frequency 
(RF)-curvature-related energy spread is substantial, the frequency of the laser pulse 
produced will also be modulated \cite{PRL}: 
\begin{equation}
f_{\rm RF} (\xi; \lambda_{\rm RF}) \approx 
f_0 \left({{1}\over{2}} +{{\lambda_{\rm RF}}\over{{8\pi \xi}}} 
\sin\left(4 \pi \xi/\lambda_{\rm RF}\right)
\right), \nonumber
\end{equation}
where $\lambda_{\rm RF}$ is the RF wavelength \cite{PRL}.
\end{enumerate}
%
%
\begin{figure*}[htb]
\includegraphics[height=1.51in]{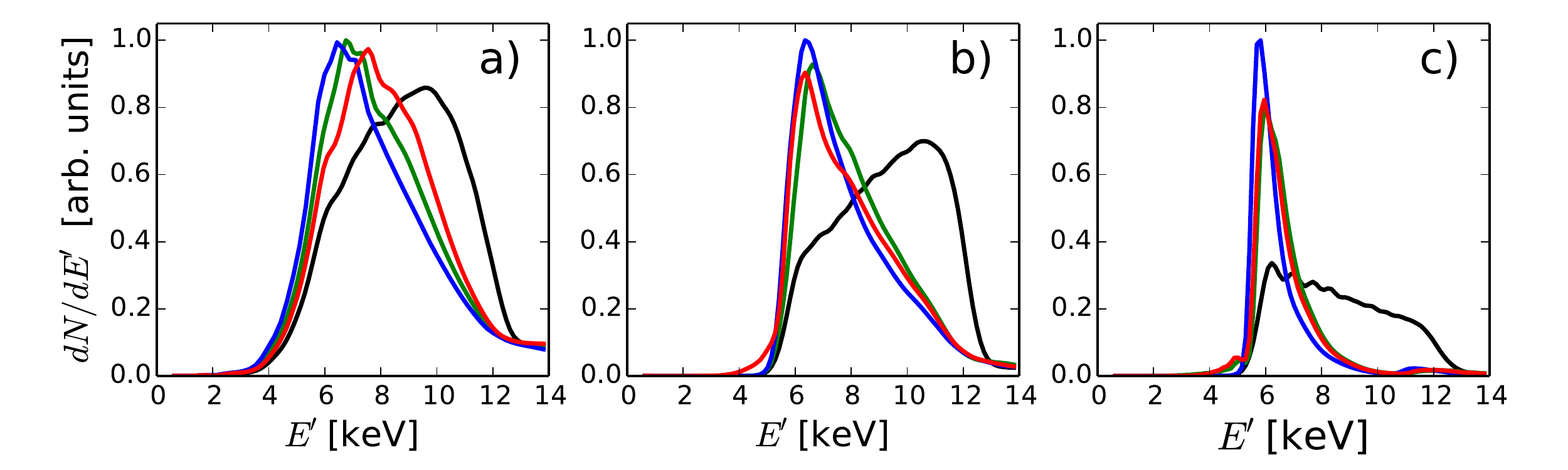}
\includegraphics[height=1.51in]{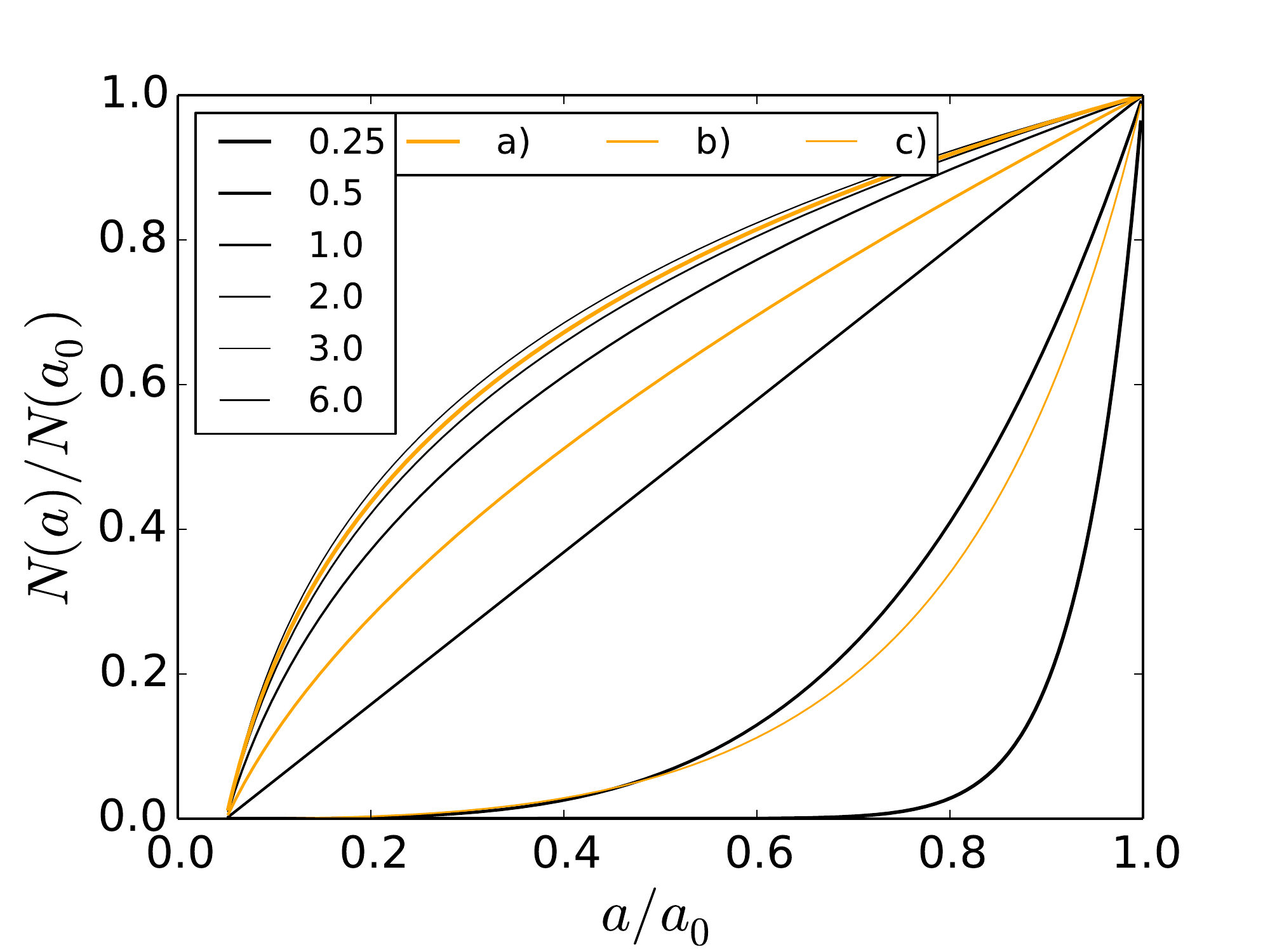}
\vskip-5pt
\caption{\small{
Panels a)-c): Simulation  of the Dresden experiment \cite{ketal2018} for $a_0 = 1.6$ using SENSE
without FM (black lines) and with FM: optimal 1D plane-wave 
$f_{\rm 1D}$ \cite{PRL} (blue lines), optimal 3D laser pulse $f_{\rm 3D}$ \cite{metal2018}
(green lines) and RF FM \cite{PRL} $f_{\rm RF}$ (red lines). The transverse electron beam size 
is nominal in a), reduced by $\sqrt{10}$ in b) and reduced by 10 in c).
Far right: Distribution of field strength values $a$ seen by the electron beam for various 
transverse ratios ($r_x=r_y=0.25,0.5,1,2,3,6$) in black. 
Shown in orange are the distributions corresponding to Panel a) with $r_x=3$, $r_y=6$, 
Panel b) with $r_x=1$, $r_y=2$, and Panel c) with $r_x=0.3$, $r_y=0.6$.
Lower cutoff of $a>a_{\rm min}=0.05$ is imposed on the distribution. 
}}
\label{fig_chirp}
\end{figure*}
%
$f_0$ is a normalization constant such that $f_0 \omega_0$ is the laser frequency at the center 
of the pulse. The normalization constant $f_0$ shifts the scattered energy spectrum by a factor 
$f_0$, without changing its shape. In the original derivation of the optimal FM for a 1D 
plane-wave \cite{PRL}, $f_0 = 1/(1+a_0^2/2)$ was adopted, such that $f_{\rm 1D}(0) = 1$, 
while consequent studies used $f_0$ such that $f(\pm\infty)=1$ \cite{retal2016,tk2016}. 
In the original derivation of the optimal FM for a 3D laser pulse \cite{metal2018}, 
$f_0$ such that $f(\pm\infty)=1$ was applied. 
In the simulations reported here, $f_0$ was chosen to make $f(0)=1$ for all three
FMs: numerically computed for $f_{\rm 3D}$, $f_0 = 1/(1+a_0^2/2)$ for $f_{\rm 1D}$
and $f_0 = 2/3$ for $f_{\rm RF}$.

All three FMs are one-parameter functions. 
The peak spectral density is maximized by carrying out a systematic search over their 
respective parameters using a genetic algorithm \cite{hetal2013,PRL}.

The distribution of laser field strength $a$ that a gaussian-distributed electron beam ``sees'' as it 
passes through the center of a gaussian laser pulse is quantified by a cumulative distribution  
\begin{equation} 
{{N (a)}\over{N(a_0)}} =  
1 - {{4}\over{\pi}} \int_{0}^{h(a)} dy 
\int_{0}^{g(a,y)}
dx~\exp\left( -x^2 - y^2\right),
\end{equation}
where 
$g(a,y) = \sqrt{\left[\log\left({{a}/{a_0}}\right) + r_y^2 y^2\right]/r_x^2}$, 
$h(a) = \sqrt{-\log\left({{a}/{a_0}}\right)/r_y^2}$.
$r_x \equiv {{\sigma_x}/{\sigma_{l,x}}}$, $r_y \equiv {{\sigma_y}/{\sigma_{l,y}}}$ are the
ratios of transverse sizes of the two beams.
Typical distributions are shown in Fig.~\ref{fig_chirp} d). 

The smaller the ratios $r_x,r_y$, the larger the transverse size of the laser pulse in comparison 
to that of the electron beam, and the more peaked the distribution $N(a)/N(a_0)$ around $a=a_0$. 
Vanishing ratios $r_x, r_y$ lead to the 1D plane-wave model in which all electrons 
experience the same strength of the laser field $a=a_0$. 
The closer the beam sizes are to this 1D plane-wave limit, the more effective the FM.
This is shown in Fig.~\ref{fig_chirp}.

\begin{figure}[htb]
\vskip-10pt
\includegraphics[width=3.45in]{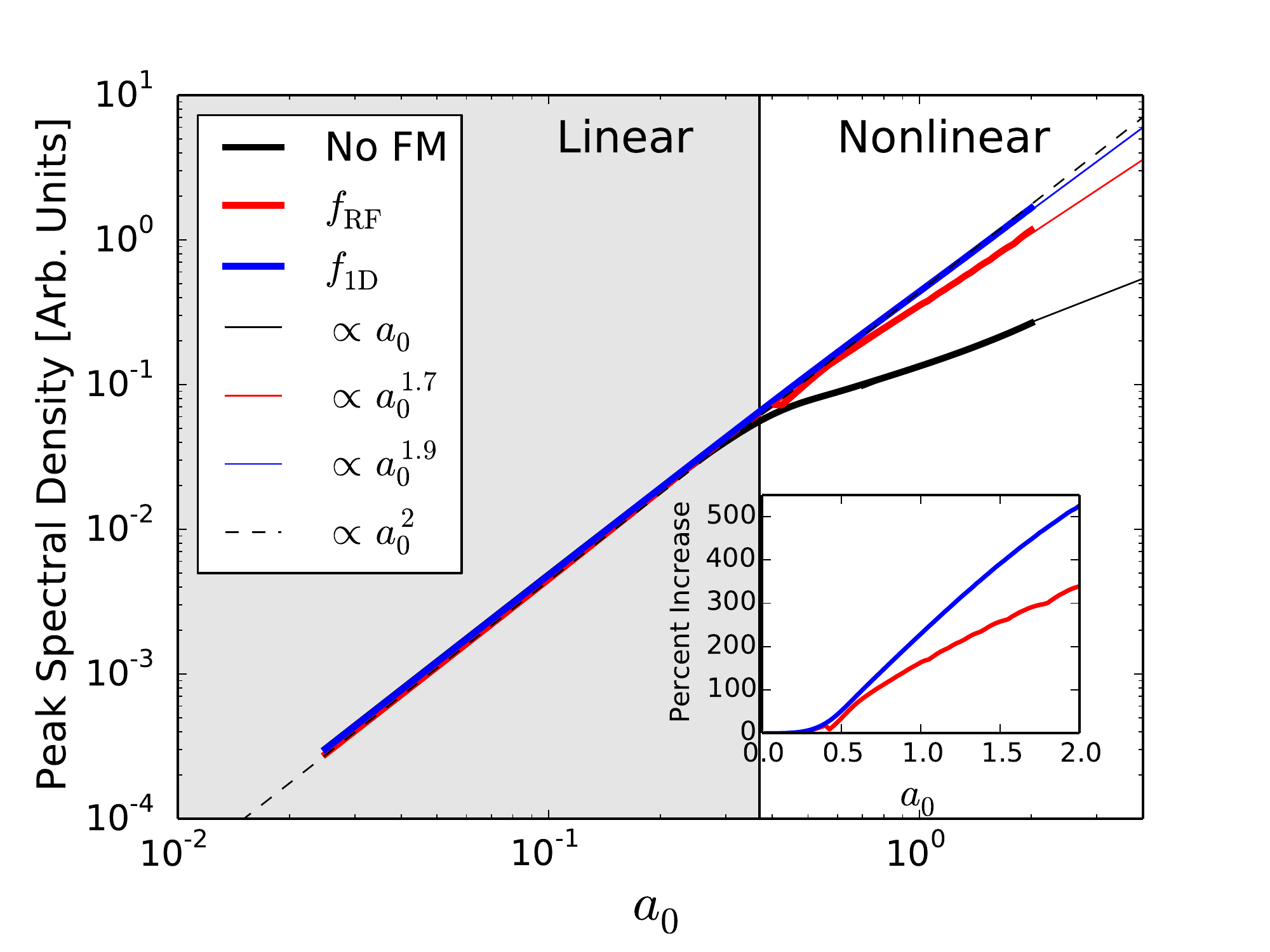}
\vskip-10pt
\caption{\small{
The peak spectral density of back-scattered radiation for an on-axis electron
passing through the laser as in the Dresden experiment \cite{ketal2018}, as a function of the 
laser field strength $a_0$, with 1D FM (red lines), RF FM (blue lines) and without FM 
(black lines). This is a 1D plane-wave limit in which FM is most effective. 
The inset shows the percent increase due to chirping.
}}
\label{fig_onaxis}
\end{figure}

Reducing the transverse size of the electron beam relative to the laser pulse---approaching the 
1D plane-wave approximation in which most electrons ``see'' a laser field whose strength is 
narrowly distributed near the maximum value of $a_0$---makes FM more efficient in increasing 
the peak spectral density of the scattered radiation. The increase due to FM depends on the relative 
sizes of the two beams, and can easily exceed 100\% for electron beams that are half the
transverse size of the laser pulse or smaller. It is also more pronounced at larger value of the 
laser field strength, as can be seen in Fig.~\ref{fig_onaxis}. 
The scattered energy at which the peak spectral density occurs can be controlled by the normalization 
constant $f_0$ of the FM which only shifts the spectrum and does not change its shape. 
For the electron beams that are transversally small when compared to the laser pulse, when 
FM is most effective, the peak of the spectrum will be located just beyond 
$4 f_0 \gamma^2 E_{\rm laser}/(1+a_0^2/2)$.

The increase of the peak spectral density for the laser pulse without FM exhibits quadratic
dependence on the laser field strength $a_0$ in the linear regime, 
but only linear in the non-linear regime (Fig.~\ref{fig_onaxis}). 
All three FMs allow the dependence of the peak spectral density on the laser field to remain 
nearly quadratic throughout the non-linear regime, thereby substantially improving the 
return on investment in increasing laser intensity.

After comparing the efficiency of the three FM functional forms---optimal 3D laser pulse 
$f_{\rm 3D}$ \cite{metal2018}, optimal 1D plane-wave $f_{\rm 1D}$ \cite{PRL} and the 
RF-induced $f_{\rm RF}$ \cite{PRL}---we find that they all are within about 20\% of 
each other, with $f_{\rm 1D}$ performing best. 
$f_{\rm RF}$ FM, the only form of the three attainable in the lab, can lead to a substantial 
increase in the peak spectral density---exceeding a factor of two for small electron beams at 
large strengths of the laser field parameter.
This FM should be used in any future experiments involving laser chirping. 
%

\section{Conclusion}
Our new code SENSE is in excellent agreement 
with the established code CAIN, and over which it offers several crucial 
advantages: (i) superior accuracy; (ii) better efficiency in cases marred by poor statistics;
(iii) arbitrary shape of the laser pulse; and 
(iv) ability to model an arbitrary FM. 

The exceptional level of accuracy of SENSE allows us to combine it with
a multidimensional non-linear optimization tool, such as a genetic algorithm, and use it 
as both a diagnostic and an optimization tool.
The set of parameter values (emittance and the energy spread of the electron beam; 
size of the laser pulse and possibly others) which minimizes the rms difference between
the experiment and simulations pinpoints their actual experimental values.
Similarly, this optimization tool can be used to find a set of 
parameter values which maximize the peak spectral density or minimize the radiation
bandwidth, thereby improving the performance of the ICS. 

The remarkable agreement between experiments and our new code SENSE strongly suggests
that the underlying model correctly captures the relevant physics.
This translates into confidence that SENSE can accurately describe physical behavior
of collisions between electron beams and chirped laser pulses, a scenario which is yet to
be tested experimentally. Simulations with SENSE strongly suggest that judiciously 
chirping the laser pulse substantially increases the spectral density. The increase
depends on the strength of the laser field, the relative transverse sizes of the 
two beams and the form of the FM function. While for the current parameters
in the Dresden experiment the returns due to chirping would be modest ($\approx 20\%$
for $a_0=1.6$), reducing the transverse size 
of the electron beam by a factor of ten would yield a three-fold increase in the peak spectral density.

\acknowledgments
We are grateful to Mohammed Zubair for his insight and consultation into computational methods.
This paper is authored by Jefferson Science Associates, LLC under U.S.~DOE Contract 
No.~DE-AC05-06OR23177. B.T. acknowledges the support from the U.S.~National 
Science Foundation award No.~1535641.

\end{document}